\documentclass[11pt,a4paper]{article}

\usepackage{amsmath,amstext,amsfonts,amsxtra,amssymb,latexsym,bm}
\usepackage{graphics,color}
\usepackage{psfrag,subfigure}
\usepackage[pdftex]{graphicx}

\newcommand{\bs}[1]{{\mathbf{#1}}}

\begin{document}

\title{\textbf{Efficient algorithms for robust recovery of images from compressed data} }

\author{Duc-Son Pham$\dagger$ and Svetha Venkatesh$\ddagger$\\	
	\small $\dagger$ Department of Computing, Curtin University, Western Australia \\	
	\small $\ddagger$ Center for Pattern Recognition and Data Analytics (PRaDA), \\ 
	\small School of Information Technology, Deakin University, Geelong, Victoria \\
	\small Email: \texttt{dspham@ieee.org} 
	}

\maketitle

\newtheorem{assumption}{\bf Assumption}
\newtheorem{lemma}{\bf Lemma}
\newtheorem{definition}{Definition}
\newtheorem{theorem}{\bf Theorem}
\newtheorem{proposition}{\bf Proposition}
\newtheorem{corollary}{\bf Corollary}
\newtheorem{observation}{\bf Observation}
\newenvironment{proof}{{\it Proof.}}

\begin{abstract}
Compressed sensing (CS) is an important theory for sub-Nyquist sampling and recovery of compressible data. Recently, it has been extended by Pham and Venkatesh \cite{pham2012improved} to cope with the case where corruption to the CS data is modeled as impulsive noise. The new formulation, termed as robust CS, combines robust statistics and CS into a single framework to suppress outliers in the CS recovery. To solve the newly formulated robust CS problem, Pham and Venkatesh suggested a scheme that iteratively solves a number of CS problems, the solutions from which converge to the true robust compressed sensing solution. However, this scheme is rather inefficient as it has to use existing CS solvers as a proxy. To overcome limitation with the original robust CS algorithm, we propose to solve the robust CS problem directly in this paper and drive more computationally efficient algorithms by following latest advances in large-scale convex optimization for non-smooth regularization. Furthermore, we also extend the robust CS formulation to various settings, including additional affine constraints, $\ell_1$-norm loss function, mixed-norm regularization, and multi-tasking, so as to further improve robust CS. We also derive simple but effective algorithms to solve these extensions. We demonstrate that the new algorithms provide much better computational advantage over the original robust CS formulation, and effectively solve more sophisticated extensions where the original methods simply cannot. We demonstrate the usefulness of the extensions on several CS imaging tasks.
\end{abstract}


\section{Introduction}
Compressed sensing (CS) \cite{Candes_Tao06}, \cite{Donoho06} is a powerful sub-Nyquist sampling theory for the acquisition and recovery of sparse signals, that has received special attention in signal and image processing as well as other related fields such as statistics and computer science. The CS theory states that if the unknown signal is inherently sparse, then it is possible to acquire and reconstruct signal (by solving a convex optimization problem) with a much lower number of measurements that would be otherwise needed under the existing Nyquist sampling scheme. In image processing, the CS theory is particularly relevant in several applications, such as magnetic resonant imaging (MRI) \cite{Lustig_Donoho_Pauly07} or hyper-spectral imaging \cite{chan2008terahertz,guo2009l1}, where acquisition time and/or sensing hardware cost play a significant role. Also, the sparsity assumption typically holds due to, for example, inherent wavelet structure in images \cite{rao2011convex}.

In recent years,  the CS literature has seen seen significant advances in both theory \cite{baraniuk2010model,candes2011probabilistic,donoho2012sensitivity,huang2010benefit} and applications \cite{eldar2009compressed,Haupt_Nowak06,herman2009high,mishali2009blind,provost2009application,vaswani2008kalman,zibulevsky2010l1} (many of which are collected in the CS repository\footnote{\url{http://dsp.rice.edu/cs}}). There are also a variety of specialized solvers for the CS recovery problem, which are developed from different angles, such as pursuit algorithms \cite{Dai_Milenkovic09}, \cite{Mallat_Zhang93}, \cite{Needell_Tropp09}, optimization algorithms \cite{Hale_Yin_Zhang07,Kim_etal07}, a complexity regularization algorithm \cite{Haupt_Nowak06b}, and Bayesian methods \cite{Ji_Xue_Carin08}.

In this work, we focus on a particular aspect of CS recovery, wherein the emphasis is on robustness. This is originally raised by Pham and Venkatesh \cite{pham2012improved}. They recognize that existing CS recovery schemes can be statistically inefficient when the corruption of CS measurements is modeled as impulsive noise. Such impulsive corruption can occur due to bit errors in transmission, malfunctioning pixels, faulty memory locations \cite{Chan_Ho_Nikolova05}, and buffer overflow \cite{Hashimoto05}, and has been raised in many image processing works \cite{Bar_etal07,Civicioglu07,Windyga01}. To address this problem, Pham and Venkatesh \cite{pham2012improved} proposed a new formulation, known as robust CS, which combines traditional robust statistics \cite{Huber81} and existing CS into a single framework to effectively suppress outliers in the recovery. Whilst the focus of \cite{pham2012improved} is on the theoretical justification of the new formulation, they also suggested a provably convergent algorithm to solve their robust CS formulation. This majorization minimization (MM) algorithm finds the robust CS solution by iteratively solving a number of CS problems, the solutions from which converge to the true solution. However, this is not computationally efficient because each iteration involves a full CS recovery, which is always iterative in nature. 

To overcome the computational limitation of the original robust CS algorithm proposed in \cite{pham2012improved}, we propose two new algorithms that directly solve the robust CS formulation. They both have only one main loop and iteratively majorize the original robust CS objective function. One algorithm is adapted from the fast iterative shrinkage thresholding (FISTA) framework developed by Beck and Teboulle \cite{Beck_Teboulle09}, which shares the same spirit as an unpublished work of Nesterov \cite{nesterov2007gradient}. The other algorithm is based on a framework known as alternating direction method of multipliers (ADMM) \cite{boyd2011distributed}. Even though the original FISTA scheme was derived for the original CS problem, it can be used for robust CS. Our contribution is a theoretical result that allows one to compute the Lipchitz constant for the application of FISTA. Additionally, we also derive a generalized ADMM algorithm for solving the robust CS formulation efficiently, which differs from the FISTA algorithm in that operator splitting and approximation updates are used. This results in a method that has same update complexity as FISTA, but is more flexible to extend.

Furthermore, we also extend robust CS in a number of directions, including additional affine constraints, $\ell_1$-norm loss function, mixed-norm regularization, and multi-tasking. We show that the ADMM is a powerful optimization framework for the robust CS problem as it can be modified or generalized to cope with these extensions, where often other CS techniques, including FISTA, find impossible to do so. We show that the derived algorithms are simple to implement, provably convergent under the ADMM theory, and that they effectively solve complex robust CS formulations.

The paper is organized follows. Section II gives some background on robust CS, whilst Section III describes the FISTA and ADMM algorithms for solving the robust CS formulation. Section IV presents four extensions of the robust CS formulation and derive computationally efficient algorithms for solving them. Section V contains numerical experiments to demonstrate the computational efficiency of the proposed algorithms. Finally, Section VI concludes the paper.

All Matlab code to implement our methods described in this paper and reproduce our results is readily available at the following website \url{http://www.computing.edu.au/~dsp/code.php}.

\section{Background}
In compressed sensing (CS), one is interested in the recovery of a sparse signal $\bs{x}\in\mathbb{R}^N$ though the compressed measurement
	\begin{eqnarray}
		\label{EQU_CS_MODEL}
	 	\bs{y} = \bm{\Phi}\bs{x} + \bs{n}.
	\end{eqnarray}
Here, $\bm{\Phi}\in\mathbb{R}^{M\times N}$ is the CS matrix that represents the compressive sampling operation and $\bs{n}$ is additive noise. The CS matrix is required to some stable embedding conditions for stable recovery \cite{Candes_Romberg_Tao06}. As $M<N$ in the CS setting, the recovery of $\bs{x}$ from $\bs{y}$ is generally ill-posed. The CS theory has established that under an assumption that $\bs{x}$ is sparse, it is possible to recover $\bs{x}$ reliably from $\bs{y}$ with an error upper bounded by the noise strength. Among various approaches to solve the CS recovery problem, the optimization formulation often provides the best achievability for a given CS matrix
	\begin{eqnarray}
		\label{EQU_CS_OPT}
	 	\hat{\bs{x}} = \arg\min_{\bs{x}\in\mathbb{R}^N} \left\{ \frac{1}{2}\|\bs{y}-\bm{\Phi}\bs{x} \|_2^2 + \lambda \|\bs{x} \|_1\right\}.
	\end{eqnarray}
In the normal CS setting, the noise in (\ref{EQU_CS_MODEL}) is often considered Gaussian with bounded norm $\| \bs{n}\|_2 \leq \xi$ and thus the maximum error induced by a CS recovery is $\mathcal{O}(\| \bs{n}\|_2)$. However, Pham and Venkatesh \cite{pham2012improved} have discovered that when the noise is indeed impulsive, such a result will still hold for normal CS recovery but is rather inefficient. Thus, they propose a modification to the CS formulation, known as robust CS, to appropriately address the characteristics of the underlying additive noise. This is achieved by considering the robust loss function instead of the quadratic cost function in (\ref{EQU_CS_OPT})
	\begin{eqnarray}
		\label{EQU_ROBUST_OPT}
	 	\hat{\bs{x}} = \arg\min_{\bs{x}\in\mathbb{R}^N} \left\{ g(\bs{x}) + \lambda \|\bs{x} \|_1\right\}.
	\end{eqnarray}
Here,  $g(\bs{x})=\sum_{i=1}^{M}\rho(y_i-(\bm{\Phi}\bs{x})_i)$ and $\rho(r)$ is the Huber's penalty function (soft limiter) given as follows
	\begin{eqnarray}
			\label{EQU_ROBUST_COST}
		 	\rho(r) = \left\{ \begin{array}{lll}
				\frac{r^2}{2} & & |r| \leq k\nu^2 \\
				-\frac{k^2\nu^4}{2} + k\nu^2|r| & & |r| > k\nu^2,
				\end{array} \right.
	\end{eqnarray}
and its derivative is given by
	\begin{eqnarray}
		 	\psi(r) = \rho'(r) = \left\{ \begin{array}{lll}
				{r} & & |r| \leq k\nu^2 \\
				k\nu^2\mbox{sgn}(r) & & |r| > k\nu^2.
				\end{array} \right.
	\end{eqnarray}
The parameter $k$ of the Huber's penalty function is determined by the fraction of the outliers whilst the scale parameter $\nu$ is often estimated from some statistic of the median, such as the median of the absolute deviation (MAD). For detail, see \cite{Huber81}.

As $\rho(r)$ is quadratic or linear depending on the actual value of $r$, solving (\ref{EQU_ROBUST_OPT}) directly is not trivial. Pham and Venkatesh suggested that instead of solving (\ref{EQU_ROBUST_OPT}), a better alternative is to solve a series of the normal CS problems. The idea is to replace $g(\bs{x})$ with an approximate quadratic function at every outer iteration with the general form
	\begin{eqnarray}		
	 	l^{k}(\bs{x}) & = & (1/2) (\bs{v}^{k}-\bm{\Phi}\bs{x})^T\bs{W}(\bs{v}^{k}-\bm{\Phi}\bs{x})+C \\
		\label{EQU_LX}
				& = & (1/2) \|\bs{W}^{1/2}\bs{v}^{k}-\bs{W}^{1/2}\bm{\Phi}\bs{x} \|_2^2 + C,
	\end{eqnarray}
where  
	\begin{eqnarray}
	C = g(\hat{\bs{x}}^{k}) - (1/2)\bm{\psi}(\bs{y}-\bm{\Phi}\hat{\bs{x}}^{k})^T\bs{W}^{-1}\bm{\psi}(\bs{y}-\bm{\Phi}\hat{\bs{x}}^{k}),	 	
	\end{eqnarray}
	\begin{eqnarray}
		\bs{v}^{k} =\bs{W}^{-1}\bm{\psi}(\bs{y}-\bm{\Phi}\hat{\bs{x}}^{k}) + \bm{\Phi}\hat{\bs{x}}^{k}.	 	
	\end{eqnarray}
Pham and Venkatesh detailed two options for $\bs{W}$, which are commonly used in the robust statistics literature
	\begin{itemize}
	 \item Modified residuals (MR): $\bs{W}=\mu\bs{I}$
	 \item Iteratively reweighted: $w_{ii}=\psi(r_i^{k})/r_{i}^{k}$, $w_{ij=0}, i\neq j$.
	\end{itemize}
When using $l^{k}(\bs{x})$ as shown in (\ref{EQU_LX}) for $g(\bs{x})$ in (\ref{EQU_ROBUST_OPT}), the resultant problem is essentially a normal CS problem and thus considered solved.

Whilst the above strategy will work, it is inefficient because each outer iteration involves a full CS problem and it is known that the CS problem needs to be solved iteratively as well. Therefore, the double loops are the main computational deficiency of the above strategy. To address this limitation, we consider bypassing the inner CS step and thus there will be only one loop for the overall algorithm. There are two powerful optimization frameworks that are suitable for this purpose, which we describe next.

\section{Proposed Algorithms}

\subsection{FISTA Algorithm}
Fast iterative shrinkage thresholding (FISTA) is an optimization approach that effectively decouples the variables from the smooth loss function in the compressed sensing objective. This approach was proposed by \cite{Beck_Teboulle09}, which also shares the same philosophy as an unpublished work of \cite{nesterov2007gradient}. Technically, FISTA is a variant of majorization minimization (MM) algorithms \cite{Hunter_Lange04} and has a special choice for the quadratic majorization as well updates that involve historical points.

Consider minimizing a convex optimization of the form $\arg\min_{\bs{x}} f(\bs{x})$ where
	\begin{eqnarray}
	 		f(\bs{x}) = g(\bs{x}) + R(\bs{x}).
	\end{eqnarray}
Here, $g(\bs{x})$ is a smooth loss function, but the variables in this loss function are coupled. The core idea of FISTA is to consider a quadratic majorization of $g(\bs{x})$, denotes as $h(\bs{x})$, such that it effectively decouples the variables. If such decoupling is possible, the approximate problem is then easier to solve even when the regularization term $R(\bs{x})$ is possibly non-smooth (such as $\| \bs{x}\|_1$), because it can be decomposed into a number of univariate optimization problems whose solution is analytical.

The first trick of FISTA is to decouple the variables by considering the majorization at iteration $k$ and approximation point  $\bs{z}^k$
	\begin{eqnarray}
	 	h(\bs{x};\bs{z}^k) = g(\bs{z}^k) + \nabla g(\bs{z}^k)^T(\bs{x}-\bs{z}^k) + \frac{L}{2} \|\bs{x}-\bs{z}^k \|_2^2.
	\end{eqnarray}
Here, $\bs{z}^k$ is used as the approximation point rather than $\bs{x}^k$ as it involves historical updates of $\bs{x}^k$ by a careful choice, which is subsequently show in (\ref{EQU_FISTA_Z}). Also, $L$ is the Lipchitz constant of the gradient of the loss function $g(\bs{x})$ to ensure that $h(\bs{x})$ is a proper majorization of $g(\bs{x})$. Thus, at iteration $k$, FISTA finds $\bs{x}^k$ via
	\begin{eqnarray}
		\label{EQU_FISTA}
	 	\bs{x}^k & = & \arg\min_{\bs{x}} \left\{ \frac{L}{2}\|\bs{x} - \bs{v}^k \|_2^2  + R(\bs{x})\right\} 
	\end{eqnarray}
where $\bs{v}^k=\bs{z}^k - (1/L)\nabla g(\bs{z}^k)$. For the quadratic loss function $g(\bs{x})=\frac{1}{2}\|\bs{y}-\bm{\Phi}\bs{x} \|_2^2$, it can be shown that $L=2\lambda_{\rm max}(\bs{\Phi}^T\bs{\Phi})$, and $\bs{v}=\bs{z}^k - (1/L)(\bm{\Phi}^T\bm{\Phi}\bs{z}^k-\bs{y})$. For the $\ell_1$-norm regularization as in the case of CS, this results in 
	\begin{eqnarray}
		\label{EQU_FISTA2}
	 	\bs{x}^k & = & \arg\min_{\bs{x}} \left\{ \frac{L}{2}\|\bs{x} - \bs{v} \|_2^2  + \lambda\|\bs{x} \|_1\right\}. 
	\end{eqnarray}
This problem can be solved element-wise and its solution is
	\begin{eqnarray}
	 	\bs{x}^k & = & \mathsf{S}_{\lambda/L}(\bs{v}),
	\end{eqnarray}
where the soft-thresholding shrinkage operator is defined as	
	\begin{eqnarray}
		\label{EQU_SHRINKAGE}
	 	\mathsf{S}_{\tau}(\bs{x}) = \{ \bs{t}: t_i = \mbox{sign}(x_i)\max(|x_i|-\tau,0)\}.
	\end{eqnarray}
The second trick of FISTA is to use a clever update of the approximation point to speed up convergence
	\begin{eqnarray}
	 	t^{k+1} & = & \frac{1+\sqrt{1+4(t^k)^2}}{2} \\
		\label{EQU_FISTA_Z}
		\bs{z}^{k+1} & = & \bs{x}^k + \left(\frac{t^k-1}{t^{k+1}}\right)(\bs{x}^k-\bs{x}^{k-1}).
	\end{eqnarray}

The original FISTA framework can be readily used for robust CS case if $\bs{v}^k$ and the Lipchitz constant can be computed for the robust loss function. In case of $\bs{v}^k$, it can be easily seen that
	\begin{eqnarray}
	 	\bs{v}^k = \bs{z}^k - \frac{1}{L}\bm{\Phi}^T\bm{\psi}(\bm{\Phi}\bs{z}^{k}-\bs{y}).
	\end{eqnarray}
It remains to compute the Lipchitz constant. To do so, we rely on the following result:
\begin{lemma}
\label{LEMMA_L_SUM}
Let $f(x)$ be a smooth convex function on $\mathcal{X}$ and suppose that the domain $\mathcal{X}$ is divided into two regions $\mathcal{X}_1$ and  $\mathcal{X}_2$, such that $f(x)=g(x)$ if $x\in\mathcal{X}_1$ and $f(x)=h(x)$ if $x\in\mathcal{X}_2$, and  $\mathcal{X}_1 \cup \mathcal{X}_2 = \mathcal{X}$, and that $g(x)=h(x)$ for $x\in\mathcal{X}_1\cap\mathcal{X}_2$. Denote as $L_g$ and $L_h$ the Lipchitz constants of $g$ and $h$ respectively on the domains $\mathcal{X}_1$ and $\mathcal{X}_2$. Then the Lipchitz constant of $f$ is bounded by
		\[L_f \leq \{L_g+L_h \}. \]
\end{lemma}
The proof of this Lemma is detailed in the Appendix. The result implies that for mixed functions like the robust CS loss functions being considered, we just take the sum of Lipchitz constants over each continuous and bounded domain. The Lipchitz constant for the quadratic part is as before, i.e. $L_1 = 2\lambda_{\rm max}(\bm{\Phi}^T\bm{\Phi})$, whilst for the linear part we can split into negative and positive domain. In both cases, the Lipchitz constant is zero due to the fact that $\psi$ is a constant. Thus, the Lipchitz constant for the robust CS cost function is still $2\lambda_{\rm max}(\bm{\Phi}^T\bm{\Phi})$.

\subsection{ADMM Algorithm}
Alternating direction method of multipliers (ADMM) is a simple but powerful framework in optimization, which is suited for today's large-scale problems arising in machine learning and signal processing. The method was in fact developed a long ago before advanced computing power was available, and re-discovered many times under different perspectives. Recently, \cite{boyd2011distributed} has unified the framework in a simple and concise explanation. In either the CS or robust CS problem, the main technical challenge is that the variables are coupled through $\bm{\Phi}$ in either the quadratic or robust loss function. This makes it rather difficult when the extra constraint with non-smooth $\ell_1$ norm is introduced. In principle, the problem is easier to tackle if the variables can be decoupled, so that the problem can be solved element-wise or group-wise. Using a clever trick, known as operator splitting \cite{eckstein1992douglas}, the ADMM framework suggests to separate the regularization term from the smooth term by introducing an additional variable $\bs{z}$, which is tied to the original variable via an affine constraint:
	\begin{eqnarray}	 	
		\begin{array}{llll}
		 	\min_{\bs{x},\bs{z}} &  g(\bs{x}) + \| \bs{z}\|_1 & \mbox{s.t} & \bs{x}-\bs{z} = 0			.
		\end{array}		
	\end{eqnarray}
Here, $g(\bs{x})$ is the robust CS loss function. For this type of regularized objective function, ADMM considers the following augmented Lagrangian
	\begin{eqnarray}	
		\label{EQU_LAGRANGIAN1}
	 	\mathcal{L}(\bs{x},\bs{z},\bs{y}) = g(\bs{x}) + \lambda\|\bs{z} \|_1  + \bs{w}^T(\bs{x}-\bs{z}) + \frac{\eta}{2}\|\bs{x}-\bs{z} \|_2^2.
	\end{eqnarray}
Here, $\eta$ is the parameter associated with the augmentation $\frac{\eta}{2}\|\bs{x}-\bs{z} \|_2^2$, and this is to improve the numerical stability of the algorithm. The strategy for minimizing this augmented Lagrangian is iterative updating of the primal and dual variables. With a further normalization on the dual variable $\bs{u} = (1/\eta)\bs{w}$, it is shown \cite{boyd2011distributed} that as far as the primal and dual variables $\bs{x}$ and $\bs{z}$ are concerned
	\begin{eqnarray}
		\label{EQU_LAGRANGIAN2}
	 	\mathcal{L}(\bs{x},\bs{z};\bs{u}) = g(\bs{x}) + \lambda\|\bs{z} \|_1  +\frac{\eta}{2}\|\bs{x}-\bs{z} + \bs{u} \|_2^2 + \mbox{const}.
	\end{eqnarray}
where the constant is independent of $\bs{x}$ and $\bs{z}$ (actually $\mbox{const}=-\eta\|\bs{u}\|_2^2/2$). Note of the semi-colon, which treats $\bs{u}$ as a parameter rather than a variable when solving for other variables. Thus, the optimality point of the Lagrangian can be found by iteratively updating the variables as follows:
	\begin{eqnarray}
		\label{EQU_X_STEP}
	 	\bs{x}^{k+1} &= & \arg\min_{\bs{x}} \left\{ g(\bs{x}) + \frac{\eta}{2}\|\bs{x} - \bs{z}^k+\bs{u}^k \|_2^2 \right\}\\
		\label{EQU_Z_STEP}
		\bs{z}^{k+1} &= & \arg\min_{\bs{z}} \left\{ \lambda\|\bs{z} \|_1 + \frac{\eta}{2}\|\bs{x}^{k+1} - \bs{z}+\bs{u}^k \|_2^2 \right\} \\
		\label{EQU_U_STEP}
		\bs{u}^{k+1} & = & \bs{u}^{k} + \bs{x}^{k+1}-\bs{z}^{k+1}.
	\end{eqnarray}
We note that the update steps for $\bs{u}$ and $\bs{z}$ are straightforward. In particular, for $\bs{z}$ it is known that it is a soft-thresholding shrinkage operation
	\begin{eqnarray}
		\label{EQU_Z_FINAL}
	 	\bs{z}^{k+1} & = & \mathsf{S}_{\lambda/\eta}(\bs{x}^{k+1}+\bs{u}^k).
	\end{eqnarray}

Due to the nature of $g(\bs{x})$, there is no exact solution for (\ref{EQU_X_STEP}), and finding it always necessitates iterative algorithms. This will increase computational burden to the overall algorithm in a similar way as the previous robust CS algorithms introduced in \cite{pham2012improved}. To alleviate the computational problem, we propose to follow a novel framework, known as generalized ADMM and developed by Eckstein and Bertsekas \cite{eckstein1992douglas}. In generalized ADMM, the update steps can be solved \textit{approximately} as long as the differences between the exact and approximate solutions generate a summable sequence. When such a condition is satisfied, the generalized ADMM theory has proved that the algorithm will converge to the solution \cite[Theorem 8]{eckstein1992douglas}. 

To utilize the generalized ADMM theory, once again we adapt an MM algorithm to solve (\ref{EQU_X_STEP}), which is in the same spirit as the original robust CS \cite{pham2012improved}. In essence, this replaces $g(\bs{x})$ with a suitable quadratic majorization as discussed previously. The major difference is that we only perform the minimization of the majorization \textit{once}, as opposed to iteratively as in \cite{pham2012improved}.  Specifically, we propose to modify the update step for $\bs{x}$ in (\ref{EQU_X_STEP}) by using the quadratic approximation of $g(\bs{x})$  at iteration $k$ as $l^{k}(\bs{x})$ (shown in (\ref{EQU_LX}))
	\begin{eqnarray}
	\label{EQU_X_MM}
	 \bs{x}^{k+1} &= & \arg\min_{\bs{x}}  (1/2) \|\bs{W}^{1/2}\bs{v}^{k}-\bs{W}^{1/2}\bm{\Phi}\bs{x} \|_2^2 + \frac{\eta}{2}\|\bs{x} - \bs{z}^k+\bs{u}^k \|_2^2. 
	\end{eqnarray}
It can be easily recognized that the solution of this problem is exact
	\begin{eqnarray}
		\label{EQU_X_FINAL}
	 \bs{x}^{k+1} = (\bm{\Phi}^T\bs{W}\bm{\Phi}^T+\eta\bs{I})^{-1}(\bm{\Phi}^T\bs{W}\bs{v}^k + \eta(\bs{z}^k-\bs{u}^k)).	
	\end{eqnarray}
We note that the quadratic approximation of FISTA can also be used. However, the choice above leads to a better approximation and hence will converge to the true solution faster. It is also easily seen that for the MR choice of the quadratic approximation where $\bs{W}=\mu\bs{I}$, the matrix under inversion in (\ref{EQU_X_FINAL}) is fixed
	\begin{eqnarray}
		\label{EQU_X_FINAL2}
		 \bs{x}^{k+1} = (\mu\bm{\Phi}^T\bm{\Phi}^T+\eta\bs{I})^{-1}(\mu\bm{\Phi}^T\bs{v}^k + \eta(\bs{z}^k-\bs{u}^k)).	
	\end{eqnarray}
Hence, the inversion $(\mu\bm{\Phi}^T\bm{\Phi}^T+\eta\bs{I})^{-1}$ can be computed once and cached so that the update step in subsequent iterations can be fast. 

The generalized ADMM for the specific case being considered can be stated as follows:
\begin{theorem}
Consider an ADMM algorithm that solves the convex problem (\ref{EQU_ROBUST_OPT}) via the updates (\ref{EQU_X_FINAL2}), (\ref{EQU_Z_STEP}), (\ref{EQU_U_STEP}). Denote as $\bs{x}^{k+1}_*$ the exact solution of (\ref{EQU_X_STEP}), and as $\bs{x}^{k+1}$ the approximate of (\ref{EQU_X_STEP}) via (\ref{EQU_X_FINAL2}). If the sequence $\{\mu_{k+1}:  \mu_{k+1} = \|\bs{x}^{k+1} - \bs{x}^{k+1}_* \|_2\}$ is summable, i.e., $\sum_{k=1}^{\infty}\mu_k\leq \infty$, then the above updates will generate a sequence $\{ \bs{x}^{k+1}\}$ that converge to the true solution of (\ref{EQU_ROBUST_OPT}).	
\end{theorem}

Next, we discuss the convergence stopping condition of the proposed generalized ADMM algorithm. When the update steps are solved exactly, the existing ADMM theory \cite{boyd2011distributed} states that the penalty parameter $\eta$ affects both the primal residual (defined as $\bs{s}^{k+1}=\eta(\bs{z}^{k+1}-\bs{z}^k)$), and the primal residual (defined as $\bs{r}^{k+1}=\bs{x}^{k+1}-\bs{z}^{k+1}$) in an opposite manner: a large $\eta$ tends to generate a small primal residual and a large dual residual and vice versa. Thus, selecting the optimal penalty parameter is typically a trade-off between primal and residual residuals with an ADMM algorithm, and $\eta=1$ generally works for most cases. However, more emphasis should be made to the primal residual in the case of the proposed generalized ADMM algorithm because the update step of the primal variable $\bs{x}$ is not solved exactly. This will ensure that the approximation error in the primal variable is promptly compensated by the dual update, at the small sacrifice in convergence rate due to the residual error being slightly larger. Intensive numerical studies suggest that a value for $\eta$ of between 2 and 5 for $\eta$ works rather well in many cases. We shall examine this in more detail in the experimental section, where we use $\eta=2$. For stopping condition, we terminate the algorithm when the primal and dual variables are sufficiently small. For standard settings of absolute and relative tolerances please see \cite{boyd2011distributed}.

\section{Beyond Robust CS}
The FISTA and ADMM algorithms for robust CS presented tackle the optimization from slightly different angles. Whilst FISTA solves the problem by replacing the robust cost function with a simpler quadratic approximation that decouples the variables, the ADMM decouples the $\ell_1$ regularization norm via operator splitting. Whilst FISTA has only one approximation, ADMM involves operator splitting \textit{and} quadratic approximation at the step that updates $\bs{x}$. Thus it appears that FISTA may have a convergence advantage due to being simpler and having less tuning requirements. However, numerical experience indicates that for a given tolerance, the ADMM algorithm is actually faster than FISTA in terms of both number of iterations or computational time to reach a given tolerance. This will be illustrated further in the experimental section.


The advantage of ADMM is better realized when one needs to extend robust CS in similar ways as many extensions on the basic CS have been made in the literature. This is difficult, if not impossible, with the FISTA scheme. Next, we discuss several possible extensions that can be simply achieved with the proposed ADMM algorithm. 

\subsection{Additional Affine Constraints}
In some cases, one would like to impose additional affine constraints on the optimization problem $\bs{c}^T\bs{x}=1$. This could be of prior knowledge on the power modeling (i.e., when $\sum_{i}x_i$ is known a priori) and this could potentially improve stabilization of the CS solution. Thus, the Lagrangian (\ref{EQU_LAGRANGIAN1}) could be altered as follows
	\begin{eqnarray}
		\mathcal{L}(\bs{x},\bs{z},\bs{y}_1,y_2) & = & g(\bs{x}) + \lambda\|\bs{z} \|_1  + \bs{w}_1^T(\bs{x}-\bs{z}) + \frac{\eta_1}{2}\|\bs{x}-\bs{z} \|_2^2  \nonumber \\
			& & + w_2(\bs{c}^T\bs{x}-1) + \frac{\eta_2}{2}\|\bs{c}^T\bs{x}-1\|_2^2.
	\end{eqnarray}
Here, $\bs{w}_1$ and $w_2$ are the dual variables for the equality constraints. Again, by scaling the dual variables $\bs{u}_1=\bs{w}_1/\eta_1$ and $u_2 = w_2/\eta_2$ we obtain
	\begin{eqnarray}
		\mathcal{L}(\bs{x},\bs{z};\bs{u}_1,u_2) & = & g(\bs{x}) + \lambda\|\bs{z} \|_1  + \frac{\eta_1}{2}\|\bs{x}-\bs{z}+\bs{u}_1 \|_2^2  + \frac{\eta_2}{2}\|\bs{c}^T\bs{x}-1+u_2\|_2^2 + \mbox{const}.
	\end{eqnarray}
Thus, the ADMM update step for $\bs{x}$ is the solution of the problem
	\begin{eqnarray}
	 	\bs{x}^{k+1}  & = &  \arg\min_{\bs{x}} \left\{ g(\bs{x}) +  \frac{\eta_1}{2}\|\bs{x}-\bs{z}^k+\bs{u}_1^k \|_2^2  + \frac{\eta_2}{2}\|\bs{c}^T\bs{x}-1+u_2^k\|_2^2 \right\}.
	\end{eqnarray}
Once again, if this step is to be solved approximately using a quadratic majorization with $\bs{W}=\mu\bs{I}$ as discussed previously then it can be shown that
	\begin{eqnarray}
	 	\bs{x}^{k+1} = \bs{H}(\mu\bm{\Phi}^T\bs{v}^k + \eta_1(\bs{z}^k-\bs{u}^k)+\eta_2(1-u_2^k)\bs{c}),
	\end{eqnarray}
where $\bs{H} = (\mu\bm{\Phi}^T\bs{W}\bm{\Phi}+\eta_1\bs{I}+\eta_2\bs{c}\bs{c}^T)^{-1}$.
		
It can be shown that the updates step for $\bs{z}$ remains the same as (\ref{EQU_Z_FINAL}) except that $\bs{u}$ and $\eta$ are replaced with $\bs{u}_1$ and $\eta_1$ respectively. Finally, the updates of the dual variables are
	\begin{eqnarray}
	 	\bs{u}_1^{k+1} & = & \bs{u}_1^k + \bs{x}^{k+1}-\bs{z}^{k+1}, \\
		u_2^{k+1} & = & u_2^k + \bs{c}^T\bs{x}^{k+1}-1.
	\end{eqnarray}
Just like the basic ADMM algorithm, convergence is determined when both the primal and dual residuals are sufficiently small. Whilst the dual residual is as before, i.e., $\bs{s}=\eta_1(\bs{z}^{k+1}-\bs{z}^k)$, there are effectively two residual vectors $\bs{r}_1^k=\bs{x}^{k}-\bs{z}^{k}$ and $r_2^k = \bs{c}^T\bs{x}^{k}-1$. Depending on the desired accuracy requirement of a particular application, the stopping criterion can be determined accordingly (see \cite[p.19]{boyd2011distributed}).

\subsection{Mixed-Norm Regularization}

In certain situations, one may wish to impose $\ell_2$ regularization on the solution of the recovery. Such a motivation may arise from the fact that the absolute sparse model may not be realistic, and thus it is more desirable to consider
	\begin{eqnarray}		
	 	\hat{\bs{x}} = \arg\min_{\bs{x}\in\mathbb{R}^N} \left\{ g(\bs{x}) + \lambda \|\bs{x} \|_1 + \beta \|\bs{x} \|_2^2 \right\}.
	\end{eqnarray}
	Even in the sparse case, an additional quadratic regularization with a small $\beta$ could improve numerical stability against rank deficiency of $\bm{\Phi}$. In the case of quadratic loss function, i.e., $g(\bs{x})=\frac{1}{2}\|\bs{y}-\bm{\Phi}\bs{x}\|_2^2$, this is known as the elastic-net \cite{zou2005regularization}. Thus, the proposed formulation could be interpreted as a robust version of the elastic-net. The robust CS formulation is treated a special case when $\beta=0$. 

For the original elastic-net, it is easily recognized that a simple algebra can convert it to a Lasso (or CS) form, and thus it can be solved with many efficient $\ell_1$-regularization algorithms. For the proposed robust elastic-net, it is not possible because of the loss function $g(\bs{x})$ being not quadratic. However, it is trivial to show that it is possible to modify the FISTA and generalized ADMM algorithms discussed in the previous section to cater for this additional regularization term. Indeed, this regularization term only affects the update step of $\bs{x}$. In both FISTA and generalized ADMM, the majorization is a quadratic function and thus absorbing this extra quadratic term is straightforward. For example, in the case of the FISTA algorithm, we need to solve (c.f. (\ref{EQU_FISTA})
	\begin{eqnarray}
	 	\min_{\bs{x}} \ \ (1/2)\|\bs{v}-\bs{x}\|_2^2 + (\lambda/L) \|\bs{x} \|_1 + (\beta/L)\|\bs{x}\|_2^2,
	\end{eqnarray}
which is equivalent to
	\begin{eqnarray}
	 	\min_{\bs{x}} \ \ \frac{1}{2}\left\|\frac{\bs{v}}{1+(\beta/L)}-\bs{x}\right\|_2^2 + \frac{\lambda}{L+\beta}\|\bs{x} \|_1, 
	\end{eqnarray}
which is of the same form and this induces the soft-thresholding shrinkage operation. Likewise, in the case of the generalized ADMM algorithm, we need to to solve (c.f (\ref{EQU_X_MM}))
	\begin{eqnarray}		
	 \bs{x}^{k+1} &= & \arg\min_{\bs{x}}  (1/2) \|\bs{W}^{1/2}\bs{v}^{k}-\bs{W}^{1/2}\bm{\Phi}\bs{x} \|_2^2 + \frac{\eta}{2}\|\bs{x} - \bs{z}^k+\bs{u}^k \|_2^2 + \| \beta\|_2^2,
	\end{eqnarray}
and thus this has only a slight modification compared with (\ref{EQU_X_FINAL})
	\begin{eqnarray}		
		 \bs{x}^{k+1} = (\mu\bm{\Phi}^T\bm{\Phi}^T+(\eta+\beta)\bs{I})^{-1}(\mu\bm{\Phi}^T\bs{v}^k + \eta(\bs{z}^k-\bs{u}^k)).	
	\end{eqnarray}
Thus, extension to mixed-norm regularization is straightforward of the proposed ADMM algorithm.

\subsection{$\ell_1$ Loss Function}
In the original robust CS paper \cite{pham2012improved}, the Huber loss is selected. This is suitable for impulsive noise being modeled as a contaminated mixture \cite{Huber81}. However, the robust CS framework is not necessarily restricted to the Huber loss function and indeed many loss functions in the robust statistics can be used to cater for different noise types. One particular interest is the $\ell_1$-norm loss function, which is optimal when the impulsive noise is modeled as a Cauchy distribution \cite{Huber81}. In this case, $g(\bs{x}) = \|\bs{y}-\bm{\Phi}\bs{x} \|_1$ and thus it is desirable to solve
	\begin{eqnarray}
	 	\hat{\bs{x}} = \arg\min_{\bs{x}\in\mathbb{R}^N} \left\{ \|\bs{y}-\bm{\Phi}\bs{x} \|_1 + \lambda \|\bs{x} \|_1\right\}.
	\end{eqnarray}
We note that the FISTA algorithm is not easily derived, because the loss function is not differentiable. 

To overcome the difficulty associated with two parts of the objective function that are both non-differentiable, we propose to apply the operator splitting mechanism of the ADMM framework twice. Specifically, we introduce  two additional variables $\bs{v}$ and $\bs{z}$ and rewrite the formulation as
	\begin{eqnarray}
	 	\arg\min_{\bs{x},\bs{v},\bs{z}} && \|\bs{v}\|_1 + \lambda \|\bs{z}\|_1  \nonumber \\
		\mbox{s.t.} && \bm{\Phi}\bs{x} -\bs{v} -\bs{y} = \bs{0} \nonumber \\
			&& \bs{x}-\bs{z} = \bs{0}.
	\end{eqnarray}
Thus, the augmented Lagrangian is
	\begin{eqnarray}
	 	\mathcal{L}(\bs{x},\bs{v},\bs{z},\bs{w}_1,\bs{w}_2) & = & \|\bs{v}\|_1 + \lambda \|\bs{z}\|_1 + \bs{w}_1^T(\bm{\Phi}\bs{x} -\bs{v} -\bs{y})  + \frac{\eta_1}{2} \| \bm{\Phi}\bs{x} -\bs{v} -\bs{y} \|_2^2 \nonumber \\
		&&  + \bs{w}_2^T(\bs{x}-\bs{z}) + \frac{\eta_2}{2}\|\bs{x}-\bs{z}\|_2^2.
	\end{eqnarray}
With the scaled dual variables $\bs{u}_1=\bs{w}_1/\eta_1$ and $\bs{u}_2=\bs{w}_2/\eta_2$, we can rewrite
	\begin{eqnarray}
	 	\mathcal{L}(\bs{x},\bs{v},\bs{z};\bs{u}_1,\bs{u}_2) & = & \|\bs{v}\|_1 + \lambda \|\bs{z}\|_1 + \frac{\eta_1}{2}\|\bm{\Phi}\bs{x}-\bs{v}-\bs{y}+\bs{u}_1\|_2^2 + \frac{\eta_2}{2} \|\bs{x}-\bs{z}+\bs{u}_2 \|_2^2 +\mbox{const}.
	\end{eqnarray}
With this form, the updates for the variables are easily computed under the ADMM principle. For $\bs{x}$, the update solves the problem
	\begin{eqnarray}
	 	\bs{x}^{k+1} & = & \arg\min_{\bs{x}} \frac{\eta_1}{2}\|\bm{\Phi}\bs{x}-\bs{v}^k-\bs{y}+\bs{u}_1^k\|_2^2 + \frac{\eta_2}{2} \|\bs{x}-\bs{z}^k+\bs{u}_2^k \|_2^2
	\end{eqnarray}
which yields the exact solution
	\begin{eqnarray}
	 	\bs{x}^{k+1} & = & (\eta_1\bm{\Phi}^T\bm{\Phi}+\eta_2\bs{I})^{-1}(\eta_1\bm{\Phi}^T(\bs{v}^k+\bs{y}-\bs{u}_1^k)  +\eta_2(\bs{z}^k-\bs{u}_2^k)).
	\end{eqnarray}
For both $\bs{v}$ and $\bs{z}$, it is easily recognized that the update steps are simple soft-thresholding  operations. For $\bs{v}$, the update step solves
	\begin{eqnarray}
	 	\bs{v}^{k+1} & = & \arg\min_{\bs{v}} \|\bs{v}\|_1 + \frac{\eta_1}{2}\|\bs{t}^k-\bs{v} \|_2^2,
	\end{eqnarray}
where $\bs{t}^k = \bm{\Phi}\bs{x}^{k+1}-\bs{y}+\bs{u}_1^k$. Likewise, for $\bs{z}$ the update step solves
	\begin{eqnarray}
	 	\bs{z}^{k+1} & = & \arg\min_{\bs{z}} \lambda\|\bs{z}\|_1 + \frac{\eta_2}{2}\|\bs{x}^{k+1}+\bs{u}_2^k-\bs{z} \|_2^2.
	\end{eqnarray}
They both have a similar form as (\ref{EQU_Z_STEP}), and thus from (\ref{EQU_Z_FINAL}) we deduce (c.f. (\ref{EQU_SHRINKAGE}))
	\begin{eqnarray}
	 	\bs{v}^{k+1} & = & \mathsf{S}_{1/\eta_1}(\bs{t}^k),  \\
		\bs{z}^{k+1} & = & \mathsf{S}_{\lambda/\eta_2}(\bs{x}^{k+1}+\bs{u}_2^k),
	\end{eqnarray}
 as the updates for $\bs{v}$ and $\bs{z}$. Finally, the dual updates are
	\begin{eqnarray}
	 	\bs{u}_1^{k+1} & = & \bs{u}_1^k + \bm{\Phi}\bs{x}^{k+1}-\bs{v}^{k+1}-\bs{y} \\
		\bs{u}_2^{k+1} & = & \bs{u}_2^k + \bs{x}^{k+1}-\bs{z}^{k+1}.
	\end{eqnarray}
The stopping criterion is when the residual vectors are sufficiently small, including $\bs{s}_1^k=\eta_1(\bs{v}^{k+1}-\bs{v}^{k})$,
$\bs{s}_2^k=\eta_2(\bs{z}^{k+1}-\bs{z}^{k})$, $\bs{r}_1^k=\bs{x}^{k}-\bs{z}^{k}$, and $\bs{r}_2^k=\bm{\Phi}\bs{x}^{k}-\bs{y}-\bs{v}^k$.

\subsection{Multi-Task Setting}
The recent literature on CS also reveals that the basic sparsity recovery scheme can be improved if one exploits further domain knowledge. Such an exploitation could be based on the constraint of the sparsity models. Extensions, such as model-based CS \cite{baraniuk2010model} and group sparsity \cite{huang2010benefit}, are key examples of the exploitation that can effectively reduce the CS requirements for a comparable recovery error when compared with conventional CS. Here, we focus on a slight variation where there are multiple CS tasks to be performed: there are multiple CS measurements $\bs{y}_i, i=1,\ldots,L$, each follows the model $\bs{y}_i=\bm{\Phi}\bs{x}_i+\bs{n}_i$.

\begin{figure}
 	\centering
	\includegraphics[width=.95\linewidth]{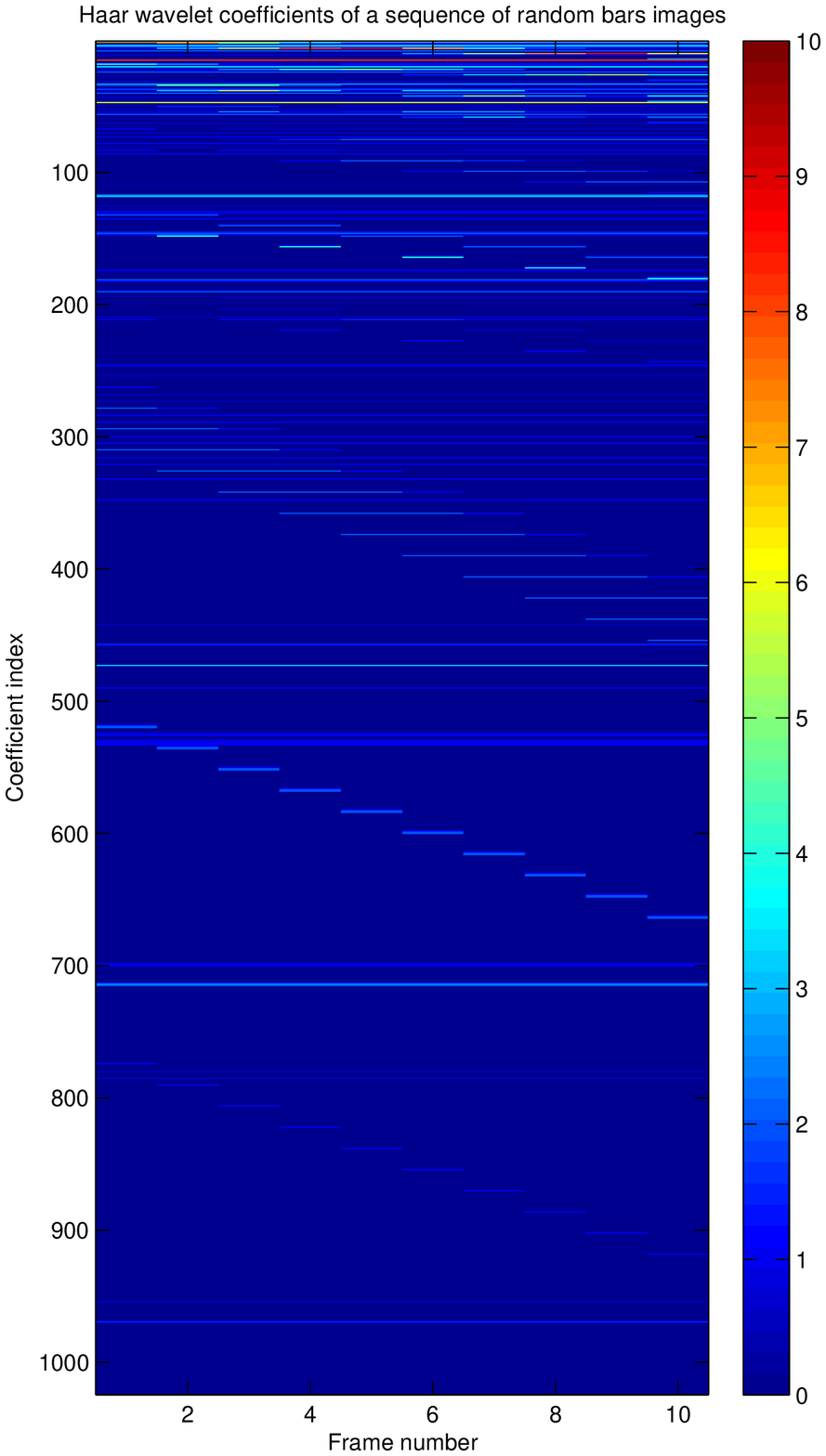}
	\caption{\label{FIG_MT_EXAMPLE} Haar wavelet coefficients - Multi random bars }
\end{figure}

In the image processing context, this could arise in, for example, compressed sensing of multiple video images. In these circumstances, there many be similarities between images. For example, moving images likely consist of relatively same large background and small moving objects. Thus, the sparse representation of these original images may have similar sparse coefficients representing the common background part (see Fig. \ref{FIG_MT_EXAMPLE} for an illustration of a sequence of random bars images used later in the experiment). For that reason, it follows from the existing results on advanced CS \cite{huang2010benefit} that exploiting the shared structure between tasks is likely to improve CS recovery compared to the case where the tasks are performed independently.

Denote as $\bs{X}=[\bs{x}_1,\ldots,\bs{x}_L]$ the collection of sparse vectors to be recovered from the tasks, and $\bs{Y}=[\bs{y}_1,\ldots,\bs{y}_L]$ the collection of CS measurements. Extending the single-task robust CS, the multi-task robust CS can be formulated as follows
	\begin{eqnarray}
	 	\hat{\bs{X}} = \arg\min_{\bs{X}\in\mathbb{R}^{N\times L}} \left\{ g(\bs{X}) + \lambda \|\bs{X}^T \|_{\ell_2/\ell_1}\right\}.
	\end{eqnarray}
Here, $g(\bs{X}) = \sum_{i=1}^{L}\rho(\bs{y}_i-\bm{\Phi}\bs{x}_i)$ and $\| \bs{A} \|_{\ell_2/\ell_1} = \sum_{i} \|\bs{a}_i\|_2$ where $\bs{a}_i$'s denote the columns of $\bs{A}$. Clearly the loss term is the same, whilst for the regularization terms, we seek sparsity along the columns of $\bs{X}$ but denseness along the rows of $\bs{X}$. This clearly reflects the prior assumption that sparse coefficients of the common parts are likely to be similar, hence the corresponding rows of $\bs{X}$ should be dense, whilst it is sparse column-wise to respect the single-task CS's assumption. When $g(\bs{X})$ is a quadratic loss function, this is a special matrix formulation of group Lasso in the statistics literature \cite{bach2008consistency,lv2011group,yuan2006model}.

We now show that it is possible to extend both the FISTA and generalized ADMM algorithms to cater for this formulation. Before doing so, we present a generalization of the soft-thresholding shrinkage operation as follows:

\begin{lemma}
The optimization problem
	\begin{eqnarray}
		\label{EQU_GS}
		\arg\min_{\bs{z}}  \ \ \left\{\lambda \| \bs{z}\|_2 + \frac{\eta}{2}\|\bs{v} -\bs{z} \|_2^2\right\}.
	\end{eqnarray}
has the solution $\bs{z} =  \bs{v} \times \max( \|\bs{v}\|_2-{\lambda}/{\eta},0)/\|\bs{v}\|_2$
\end{lemma}
This result can be proved by simple geometrical arguments. Indeed, denote $\bs{z}^*$ as the solution of (\ref{EQU_GS}), then we consider all points $\bs{z}$ such that $\|\bs{v}-\bs{z}\|_2=\|\bs{v}-\bs{z}^*\|_2=R$. It turns out that these points are lying on the ball with center at $\bs{v}$ and radius $R$. Among these points, only the point that satisfies $\bs{z}=\alpha\bs{v}$, i.e., intersection of the ball and the vector $\bs{v}$, will have minimum $\ell_2$ norm, which minimizes the second term in (\ref{EQU_GS}). Substituting this into (\ref{EQU_GS}) yields the form of the soft-thresholding shrinkage problem, for which the result is obtained after simple manipulations.

\subsubsection{FISTA algorithm.} Generalizing (\ref{EQU_FISTA}) for the multi-task settings, denote as $\bs{V}^k=[\bs{v}_1^k,\ldots,\bs{v}_L^k]$, where $\bs{v}_i^k = \bs{z}_i^k - \frac{1}{L}\bm{\Phi}^T\bm{\psi}(\bm{\Phi}\bs{z}_i^{k}-\bs{y}_i)$. Then, the update step for $\bs{X}$ solves
	\begin{eqnarray}
	 	\arg\min_{\bs{X}} \frac{1}{2} \|\bs{V}-\bs{X}\|_F^2 + \lambda \|\bs{X}^T\|_{\ell_2/\ell_1}.
	\end{eqnarray}
This problem can be written row-wise in the form of (\ref{EQU_GS}) and thus the solution is exact. Meanwhile, the update step for $\bs{Z}$ is also similar
	\begin{eqnarray}
	 	\bs{Z}^{k+1} & = & \bs{X}^k + \frac{t_k-1}{t_{k+1}}(\bs{X}^k-\bs{X}^{k-1}),
	\end{eqnarray}
with $t^{k+1}=(1+\sqrt{1+4(t^k)^2})/2$.

\subsubsection{ADMM algorithm.} We rewrite the Lagrangian for the current setting as follows
\begin{eqnarray}
 	\mathcal{L}(\bs{X},\bs{Z};\bs{U}) & = & g(\bs{X}) + \lambda \|\bs{Z}^T\|_{\ell_2/\ell_1} + \frac{\eta}{2}\|\bs{X}-\bs{Z}+\bs{U}\|_F^2 + \mbox{const}.
\end{eqnarray}
Thus, the ADMM update steps are
	\begin{eqnarray}
		\label{EQU_MX_STEP}
	 	\bs{X}^{k+1} &= & \arg\min_{\bs{X}} \left\{ g(\bs{X}) + \frac{\eta}{2}\|\bs{X} - \bs{Z}^k+\bs{U}^k \|_F^2 \right\}\\
		\label{EQU_MZ_STEP}
		\bs{Z}^{k+1} &= & \arg\min_{\bs{z}} \left\{ \lambda\|\bs{Z}^T \|_{\ell_2/\ell_1} + \frac{\eta}{2}\|\bs{X}^{k+1}+\bs{U}^k  - \bs{Z}\|_F^2 \right\} \\
		\label{EQU_MU_STEP}
		\bs{U}^{k+1} & = & \bs{U}^{k} + \bs{X}^{k+1}-\bs{Z}^{k+1}.
	\end{eqnarray}	
Like FISTA, the update step of $\bs{Z}$ can easily be decomposed row-wise, each has the form of (\ref{EQU_GS}), and thus the solution for each row of $\bs{Z}^{k+1}$ can be obtained immediately. For the update step of $\bs{X}$, again we resort to the generalized ADMM principle. That is, we approximate $g(\bs{X})$ with a quadratic loss at $\bs{X}^k$
	\begin{eqnarray}
	 	h(\bs{X};\bs{X}^k) & = & g(\bs{X}^k) + \nabla_{\bs{X}}g(\bs{X}^k)^T(\bs{X}-\bs{X}^k)  + \frac{\mu}{2}\|\bs{X}-\bs{X}^k\|_F^2.
	\end{eqnarray}
Thus, the generalized ADMM algorithm finds the update via
	\begin{eqnarray}
	 	\bs{X}^{k+1} &= & \arg\min_{\bs{X}} \left\{ \nabla_{\bs{X}}g(\bs{X}^k)^T(\bs{X}-\bs{X}^k) + \frac{\mu}{2}\|\bs{X}-\bs{X}^k\|_F^2  + \frac{\eta}{2}\|\bs{X} - \bs{Z}^k+\bs{U}^k \|_F^2 \right\},
	\end{eqnarray}
which yields the following solution
	\begin{eqnarray}
	 	\bs{X}^{k+1} = (\mu\bm{\Phi}^T\bm{\Phi}+\eta\bs{I})^{-1}(\mu\bm{\Phi}^T\bs{V}^k+\eta(\bs{Z}^k-\bs{U}^k)),
	\end{eqnarray}
where $\bs{V}^k=[\bs{v}_1^k,\ldots,\bs{v}_L^k]$, $\bs{v}_k = \frac{1}{\mu}\bm{\psi}(\bs{y}_i-\bm{\Phi}\bs{x}_i^k)+\bm{\Phi}\bs{x}_i^k$.
	
The stopping criterion is when all primal and dual residual matrices are small, they include
	\begin{eqnarray}
	 	\bs{S}^{k+1} & = & \eta (\bs{Z}^{k+1}-\bs{Z}^k) \\
		\bs{R}^{k+1} & = & \bs{X}^{k+1}-\bs{Z}^{k+1}.
	\end{eqnarray}
Like the single-task case, one should set $\eta$ sufficiently large to obtain a smooth decrease of the objective function.

\subsection{Discussion}

\textit{Further extensions.} We have presented some fundamental extensions of the CS formulation. Under the ADMM frameworks, it appears that it is possible to consider extensions based on the combination of the basics extensions presented. For example, the $\ell_1$ loss could be used with affine constraint or in multi-task setting, etc. Such extensions will be worthwhile investigation for future work.

\begin{figure}
 	\centering
	\includegraphics[width=.95\linewidth]{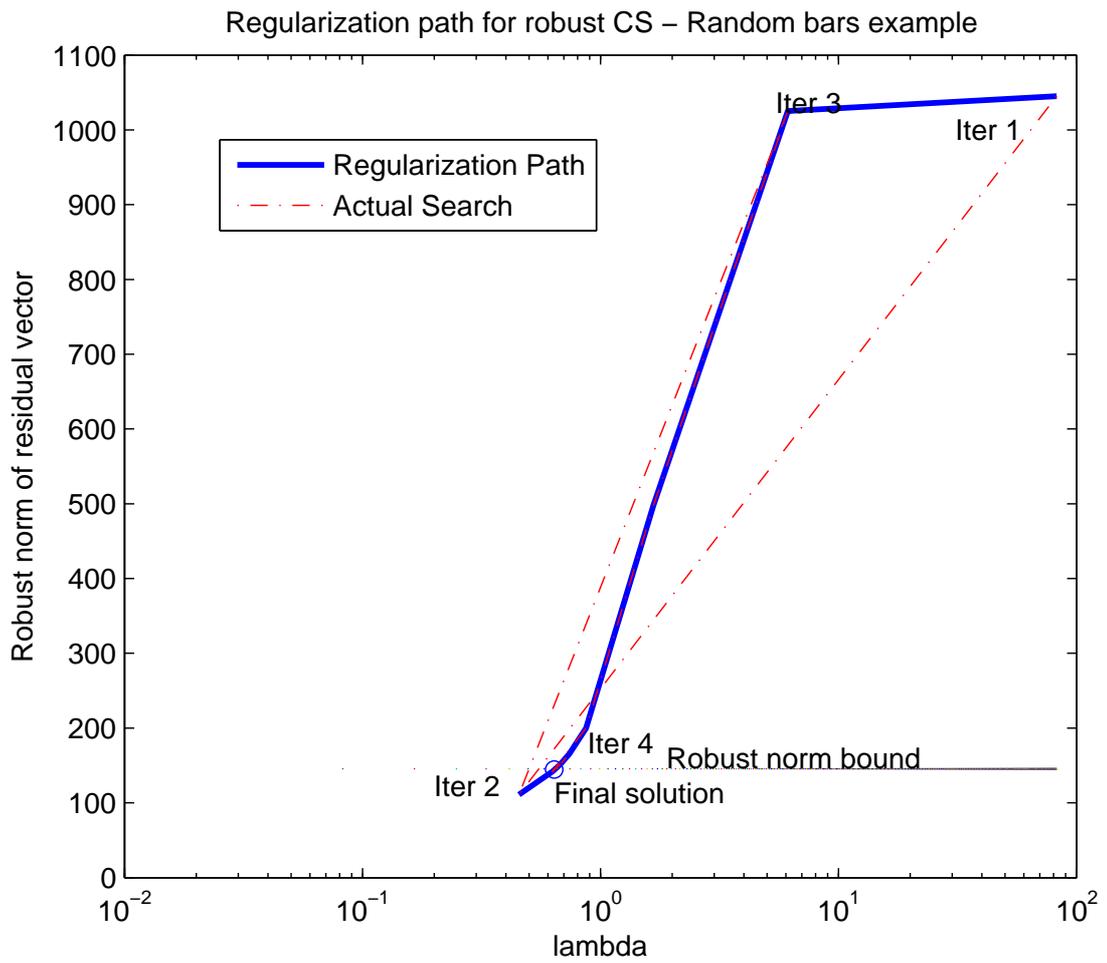}
	\caption{\label{FIG_REG_PATH}Regularization path of robust CS for random bars example}
\end{figure}

\textit{Regularization Path.} In practice, the optimal value of the regularization $\lambda$ is not known in advance, and thus one needs to select a proper value to do robust CS recovery. Such a problem is known in statistics as model selection. Typically, one needs to compute the recovery along the regularization path, and select the one which meets the $\ell_1$ norm constraint. This is discussed in detail in \cite{pham2012improved}. Essentially, some estimates of the noise statistics must be obtained in order to construct the bound on the residual $\varepsilon$. It is well-known that there exist a $\lambda_{\rm max} = \|\bm{\Phi}^T \bs{y}\|_{\infty}$ above which the solution is zero. For decreasing values of $\lambda$, the residual $\bs{r}=\bs{y}-\bm{\Phi}\hat{\bs{x}}$ will become smaller whilst the recovery becomes denser. The optimal $\lambda$ is the maximum value of $\lambda$ such that the bound constraint on the residual vector is met. In CS recovery, this happens when $\|\bs{r}\|_2^2 \leq \varepsilon$, whilst in robust CS recovery, Pham and Venkatesh \cite{pham2012improved} have suggested $\rho(\bs{r}) \leq \varepsilon$, which is a generalization of the CS selection criteria for the robust case. In our implementation, we combine a coarse grid search and a fine bi-section search to find this optimal $\lambda$ (see Fig. \ref{FIG_REG_PATH} for an illustration).  

\textit{Cholesky Factorization.} As can be seen, most update step of $\bs{x}$ in different ADMM variants involves the computation of the form $\bs{x}^{k+1} = (\mu\bm{\Phi}^T\bm{\Phi} +  \bs{Q})^{-1}\bs{q}$ where $\bs{Q}$ is an positive definite matrix. The matrix under inversion has a size of $N \times N$ and it is large in image processing application. Thus, it is inefficient to compute the inversion directly to obtain the update. A much more efficient approach is to use Cholesky decomposition to achieve the goal. It is known from linear algebra that if $\bs{H}$ is a positive definite matrix then it admits the factorization $\bs{H}=\bs{L}\bs{L}^T$ and thus $\bs{H}^{-1}\bs{q}$ can be efficiently computed by solving $\bs{L}\bs{x}_1=\bs{q}$ first, then $\bs{L}^T\bs{x}=\bs{x}_1$, which can be written as $\bs{x} = \bs{L}^T \setminus (\bs{L} \setminus \bs{q})$. For compressed sensing applications where $\bm{\Phi}$ is a fat matrix, further exploitation can be made by reducing the dimension of the matrix for Cholesky factorization. Indeed, according to the matrix inversion lemma
	\begin{eqnarray*}
	 	(\mu\bm{\Phi}^T\bm{\Phi} +  \bs{Q})^{-1} & = & \bs{Q}^{-1} - \bs{Q}^{-1}\bm{\Phi}^T \ \bs{P}^{-1}\  (\mu\bm{\Phi}\bs{Q}^{-1}),
	\end{eqnarray*}
where $\bs{P}=\bs{I} + \mu\bm{\Phi}\bs{Q}^{-1}\bm{\Phi}^T$. Suppose that the Cholesky factorization of $\bs{P}$ is $\bs{P}=\bs{L}\bs{L}^T$ then
	\begin{eqnarray*}
	 	(\mu\bm{\Phi}^T\bm{\Phi} +  \bs{Q})^{-1}\bs{q} = \bs{Q}^{-1}(\bs{q}-\bm{\Phi}^T(\bs{L}^T \setminus (\bs{L}\setminus(\mu\bm{\Phi}\bs{Q}^{-1}\bs{q})))).
	\end{eqnarray*}
We can avoid the direct inversion of $\bs{Q}$ by exploiting the fact that if $\bs{Q} = \rho_1 \bs{I} + \rho_2 \bs{c}\bs{c}^T$ then the matrix inversion lemma once again gives
	\begin{eqnarray*}
	 	\bs{Q}^{-1} = \rho_1^{-1}\bs{I}- \gamma\bs{c}\bs{c}^T,
	\end{eqnarray*}
where $\gamma = \rho_1^{-2}(\rho_2^{-1}+\rho_1^{-1}\bs{c}^T\bs{c})$. Finally, we note that this Cholesky factorization is independent of the regularization parameter $\lambda$ and thus it can be cached for the whole regularization path to reduce computation.

\section{Experiments}

\begin{figure*}
 	\centering
	\includegraphics[width=.95\linewidth]{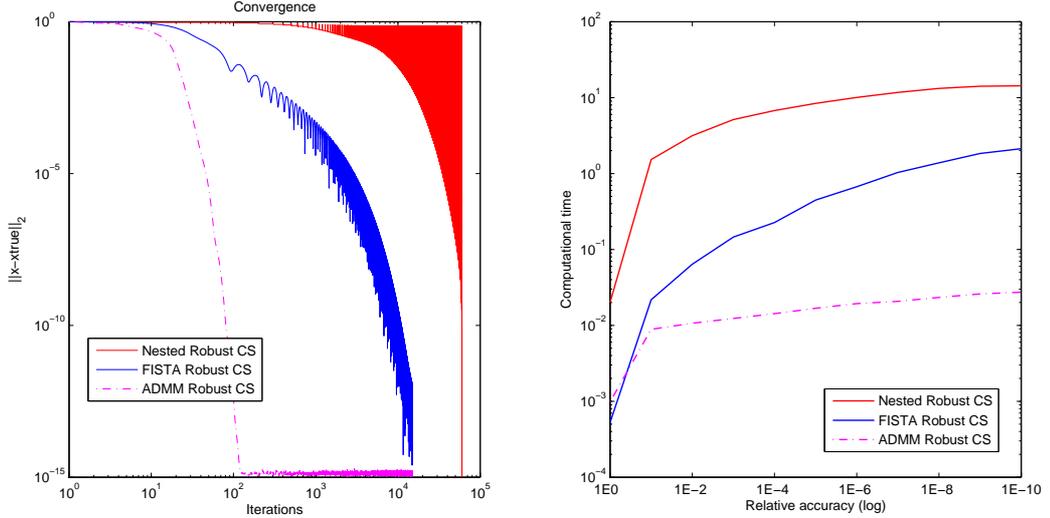}
	\caption{\label{FIG_CONV_RANDOM}Convergence property of the compared robust CS algorithms}
\end{figure*}

\begin{figure*}
 	\centering
	\includegraphics[width=.95\linewidth]{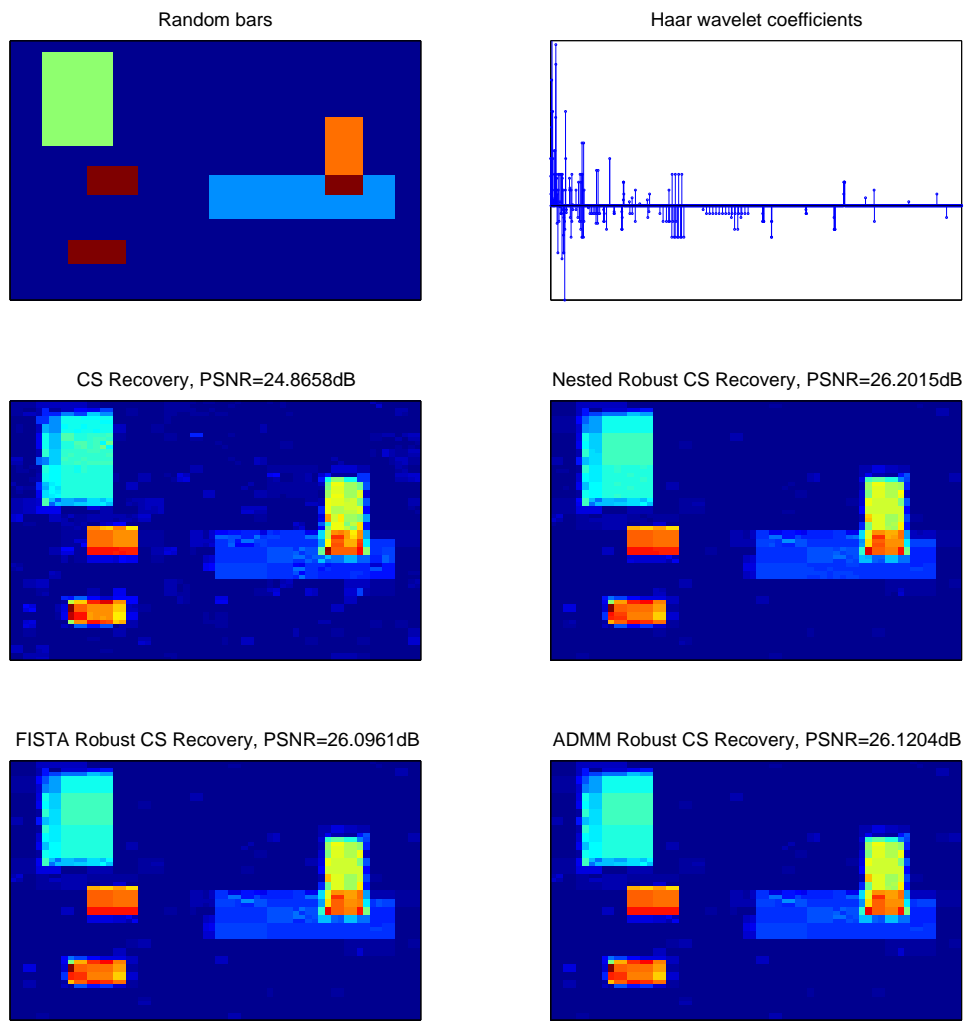}
	\caption{\label{FIG_RANDOM_BARS} Image recovery - random bars example}
\end{figure*}

\subsection{Comparison of Numerical Properties}
We examine the convergence property of the FISTA and ADMM algorithms and compare them with the previously proposed method in \cite{pham2012improved}, which we refer to as nested robust CS algorithm due to the nature of the double loops inside that algorithm. As the nested robust CS algorithm \cite{pham2012improved} is dependent on the particular CS solver being used for the inner loop, we select the ADMM implementation as the CS solver because it provides the best computational accuracy and speed. Note that Pham and Venkatesh \cite{pham2012improved} used the \texttt{l1\_ls} algorithm originally, which is known for high-accuracy but computationally expensive. However, numerical experience shows that the inner steps do not required to be solved with high accuracy. Thus, the ADMM implementation as a CS solver for the nested robust CS algorithm is better overall. In this case, it can be seen that the computational complexity per iteration (regardless of inner or outer) in all compared algorithms are approximately the same: they all involve the computation of the majorization point and the soft-thresholding shrinkage operation. 

To compare the algorithms, we examine two aspects: the error versus the iterations and the computational time taken to achieve a particular tolerance. Whilst the former indicates how fast an algorithm converges, the latter provides a much valuable insight for practical purpose. To do so, we let all algorithms run for sufficiently large number of iterations and measure the error (with respect to the true value of the robust CS solution) as iterations go on, and the computational time taken when the error reaches certain thresholds. For the ADMM-based CS solver used in the inner loop of the nested robust CS algorithm, we select the termination with relative tolerance of $10^{-2}$ and absolute tolerance of $10^{-4}$ (see \cite[p.19]{boyd2011distributed}). This allows a reasonable convergence within the inner loops. We also choose the modified residual approach for nested robust CS as it is simpler without loosing convergence advantage. All algorithms are implemented in Matlab, and roughly optimized.

We revisit the random bars example in \cite{pham2012improved} (see also Fig. \ref{FIG_RANDOM_BARS}) and the results of this study is shown in Fig. \ref{FIG_CONV_RANDOM}. In this example, the signal to noise ratio is 20dB and the impulsive noise is modeled as a two-component Gaussian mixture model where the there is 10\% contamination whose variance is $\kappa=100$ times that of the main component. Here, the left subplot shows the reduction of the error versus the iterations, whilst the right plot shows the time taken to achieve the relative accuracy from initialized zeros (as indicated by 1E0) to as small as $10^{-10}$ of the initial error (as indicated by 1E-10). We note the error profile of the nested robust CS algorithm ranges considerably due to the fact that we measure with respect to the global solution of the outer loop and that within each CS inner loop the algorithm still converges normally.

Clearly the error profile plot indicates that the ADMM algorithm offers the best convergence speed per iteration, followed by the FISTA algorithm. For example, to achieve an accuracy of $10^{-5}$ of the initial error, it only takes the ADMM algorithm less than 100 iterations, whilst the FISTA algorithm needs to spend more than 20 times, and the nested algorithm would need 200 times the number of iterations. In terms of the actual time taken to achieve a particular tolerance, the right subplot further indicates the advantage of ADMM and FISTA algorithms over the nested one. In practice, one would be interested in the tolerance of between $10^{-2}$ to $10^{-6}$, over which the ADMM and FISTA algorithms are observed to be 100 and 10 times faster than the nested algorithm respectively.

In Fig. \ref{FIG_RANDOM_BARS}, we shows the actual image recovery of all compared methods, including the CS, the nested robust CS, the FISTA robust CS, and the ADMM robust CS algorithms on this random bars example. The original random bars image is shown on the top left subplot, whilst its Haar wavelet coefficients are shown on the top right subplot. The results clearly show that all robust CS methods achieve an PSNR of about 26dB, which is 1.5dB better that that of conventional CS recovery. We note that there is a very minor different between robust CS algorithms, due to different convergence termination conditions, which is unavoidable.

\subsection{Recovery with Affine Robust CS}
Next, we examine how much improvement can be made to robust CS if the power is known. The affine robust CS formulation is slightly different to the robust CS formulation in that additional constraint $\bs{c}^T\bs{x}=1$ is imposed, and here we select $\bs{c} = 1/(\bs{1}^T\bs{x}) =  1 / \sum_i {x_i}$ and assume that $\sum_i {x_i}$ is known. 

\begin{figure*}
 	\centering
	\includegraphics[width=.95\linewidth]{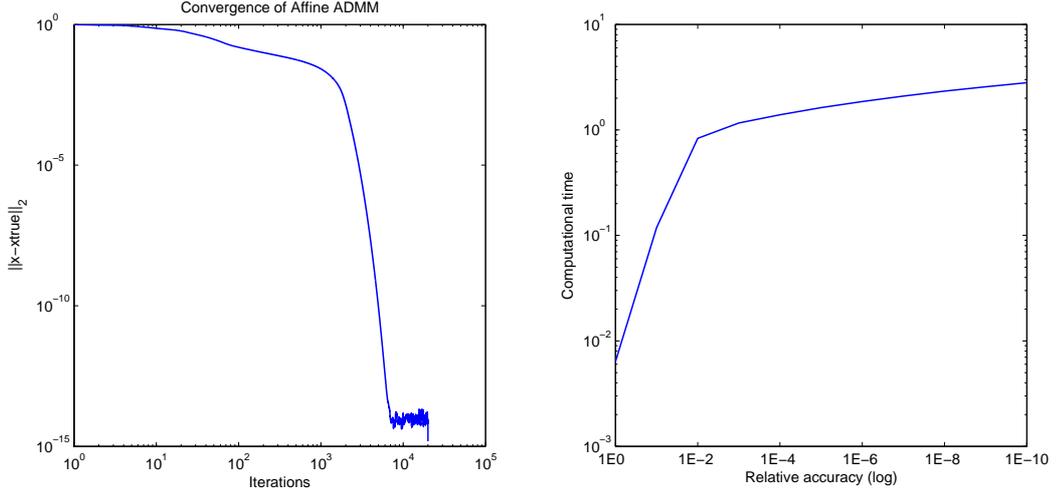}
	\caption{\label{FIG_CONV_AFFINE}Convergence behavior of affine robust CS}
\end{figure*}

First, we examine the convergence behavior of the affine ADMM robust CS algorithm to solve this formulation by revisiting the random bars example. In this case, we select $\rho_1=\rho_2=1$ and let the algorithm run over sufficient number of iterations. The results are shown in Fig. \ref{FIG_CONV_AFFINE}. Again, the left subplot shows the absolute error against the iterations whilst the right subplots indicates computational time taken to reach a particular accuracy. Compared with those of the ADMM robust CS algorithm, it can be clearly seen that the affine ADMM robust CS algorithm takes more time to reach. This is as expected because there are only minor changes to the update steps of the primal and dual variables.

\begin{figure*}
 	\centering
	\includegraphics[width=.95\linewidth]{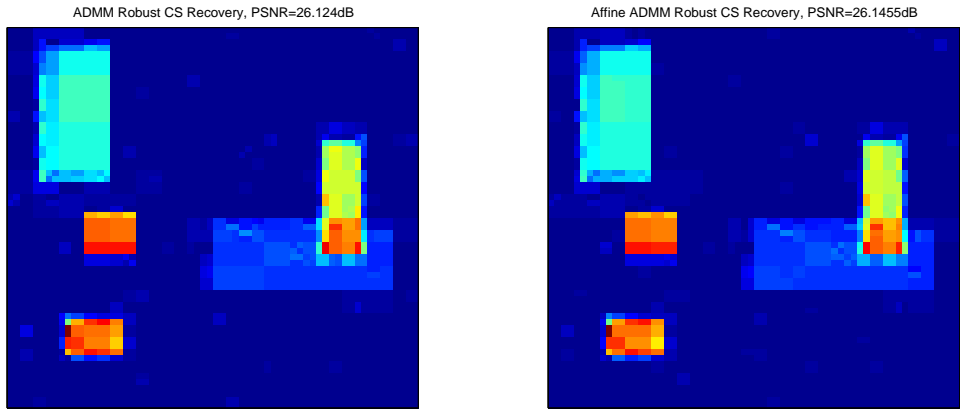}
	\caption{\label{FIG_AFFINE}Image recovery of affine robust CS formulation}
\end{figure*}

Next, we examine the actual image recovery of affine robust CS formulation. Once again, the random bars example is used and the recovered images are shown in Fig. \ref{FIG_AFFINE}. Here, we compare with the robust CS formulation via the ADMM algorithm. The result indicates that there is a slight gain in the recovery, though it is rather little. As a result, the recovered images look similar. 

\subsection{Recovery with $\ell_1$ Robust Loss Function}
Next, we demonstrate the robust CS algorithm with $\ell_1$ loss function rather than the Huber's loss function used in \cite{pham2012improved}. This is useful in situations with very impulsive corruption, where the noise is best modeled by a Cauchy distribution. To do so, we revisit the random bars example, but we use Cauchy noise instead. For the $\ell_1$ ADMM robust CS algorithm, the model selection criteria is the $\ell_1$ norm of the residual, rather than the Huber's loss function to reflect the new formulation. Other than that, all other experimental settings remain the same. 

\begin{figure*}
 	\centering
	\includegraphics[width=.95\linewidth]{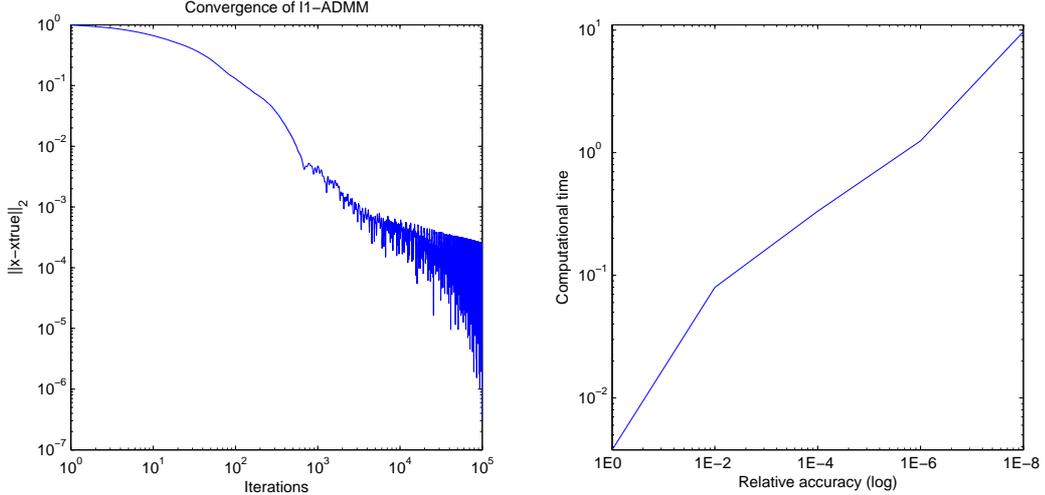}
	\caption{\label{FIG_CONV_L1_ADMM} Convergence of the $\ell_1$-loss ADMM robust CS algorithm}
\end{figure*}

First, we examine the convergence behavior of the ADMM robust CS algorithm with $\ell_1$ loss. Fig. \ref{FIG_CONV_L1_ADMM} shows the typical convergence behavior of the algorithm in terms of accuracy versus iterations (left) and computational time taken to reach certain accuracy (right). It is observed that the convergence is slower with modest accuracy as compared with the formulation using Huber's loss function. This is as expected from ADMM optimization theory due to an increasing number of variables to solve the $\ell_1$ loss formulation. Nevertheless, modest accuracy might be sufficient for many practical situations.

\begin{figure*}
 	\centering
	\includegraphics[width=.95\linewidth]{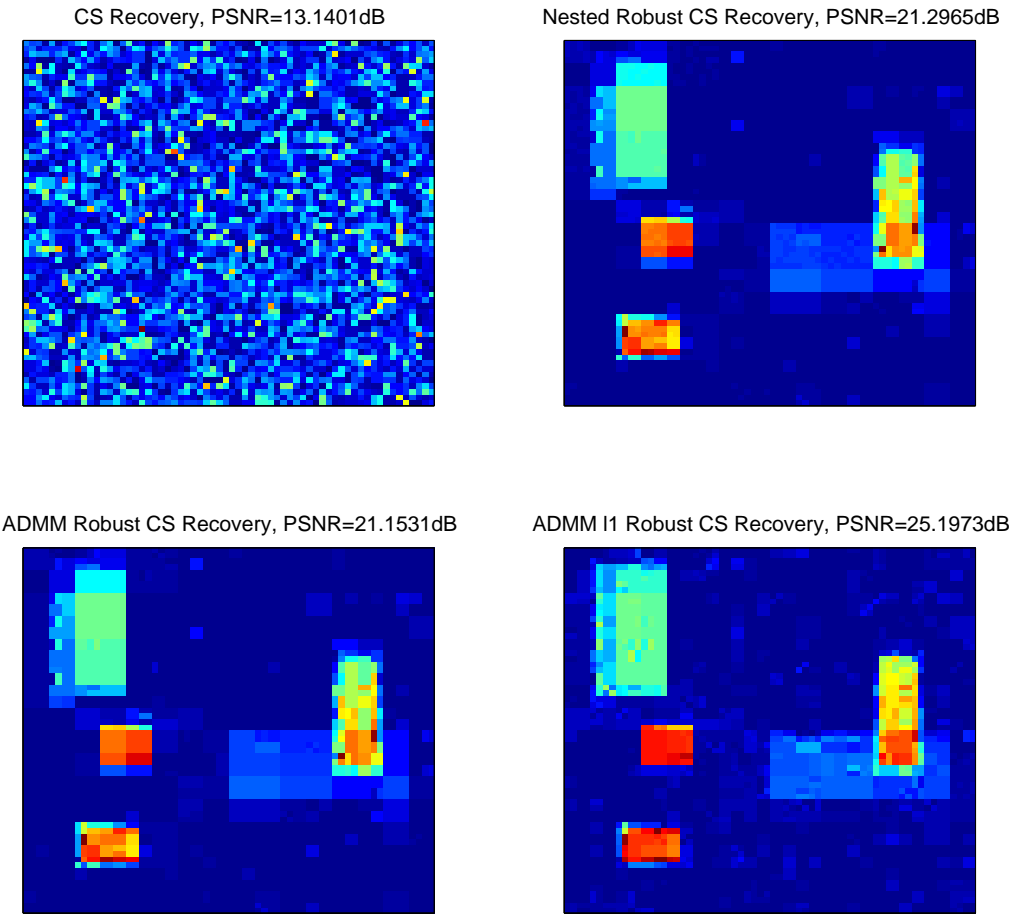}
	\caption{\label{FIG_CAUCHY} Image recovery in Cauchy noise}
\end{figure*}

Next, we examine image recovery quality in Cauchy noise. Fig. \ref{FIG_CAUCHY} shows the image recovery for CS, robust CS using nested, ADMM, and $\ell_1$-regularized ADMM algorithms respectively. Due to Cauchy noise, it is of interest to note that the CS completely fails with no meaningful pattern recovered. The other nested and ADMM algorithm still maintain reasonably recovery quality with an PSNR of around 21dB. The $\ell_1$-regularized ADMM algorithm achieves the best result with an PSNR of 25dB, a significant improvement compared with the other two robust CS algorithms. It is also noted that the computational time of the $\ell_1$-regularized ADMM algorithm is almost equal to that of the ADMM robust CS algorithm due to the fact that the update steps of the two algorithms have similar complexity. 

\begin{figure*}
 	\centering
	\includegraphics[width=.95\linewidth]{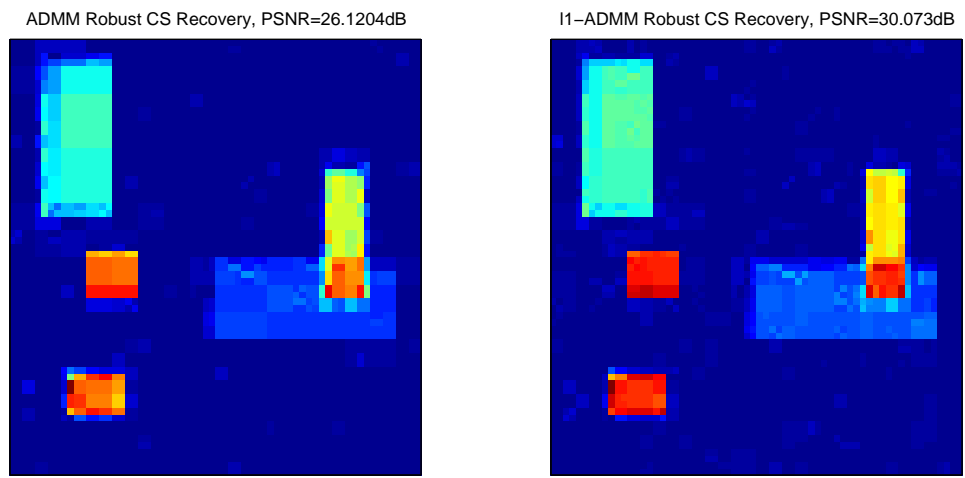}
	\caption{\label{FIG_GMM} Image recovery of robust CS with $\ell_1$ loss function in Gaussian mixture noise}
\end{figure*}

Though the $\ell_1$ loss is primarily used for noise modeling as the Cauchy distribution, it is still of interest to examine how it behaves if the noise is modeled as from the Gaussian mixture as used previously. We again revisit the settings in the previous experiment and the result is shown in Fig. \ref{FIG_GMM}. Surprisingly, the $\ell_1$-loss formulation provides a considerable PSNR gain of 4dB over the Huber's loss robust CS formulation. 

Thus, despite having less favorable convergence properties, the robust CS formulation with $\ell_1$ loss still appears a better performer for practical image recovery. 

\subsection{Recovery of A Sequence of Compressed Images}

\begin{figure*}
 	\centering
	\includegraphics[width=.95\linewidth]{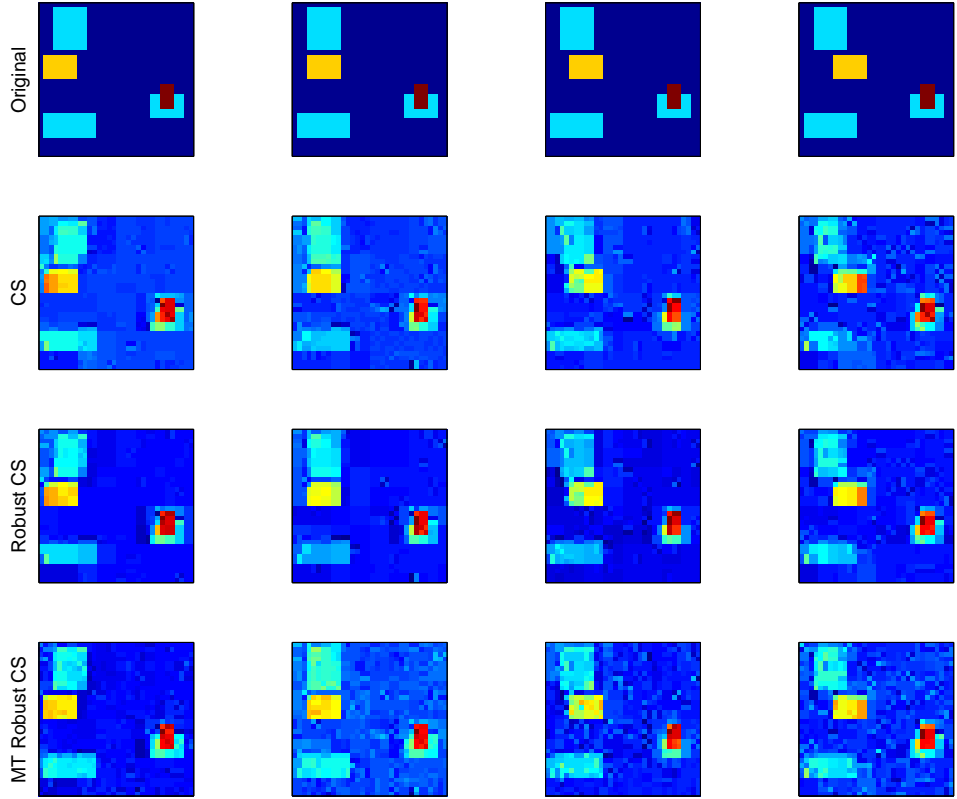}
	\caption{\label{FIG_MT_RECOVERY} Recovery of sequence of compressed images}
\end{figure*}

Finally, we demonstrate the usefulness of the multi-task robust CS formulation when a sequence of 10 compressed images corrupted by impulsive noise need to be recovered. Whilst each image in the sequence can be recovered separately, the multi-task robust CS formulation suggests that exploiting the shared structure between the tasks may provide better recovery. To do so, we consider a sequence of random bars frames shown in the top row of Fig. \ref{FIG_MT_RECOVERY}. Here, there are common static random bars and a moving block across the frames. Obviously, the wavelet coefficients for common static bars are shared between the CS tasks. Only the coefficients corresponding to the moving block distinguish between tasks. This is clearly illustrated in Fig. \ref{FIG_MT_EXAMPLE} which shows an image plot of Haar wavelet coefficients of all 10 random bars image in a sequence: the horizontal lines correspond to common coefficients.

\begin{figure*}
 	\centering
	\includegraphics[width=.8\linewidth]{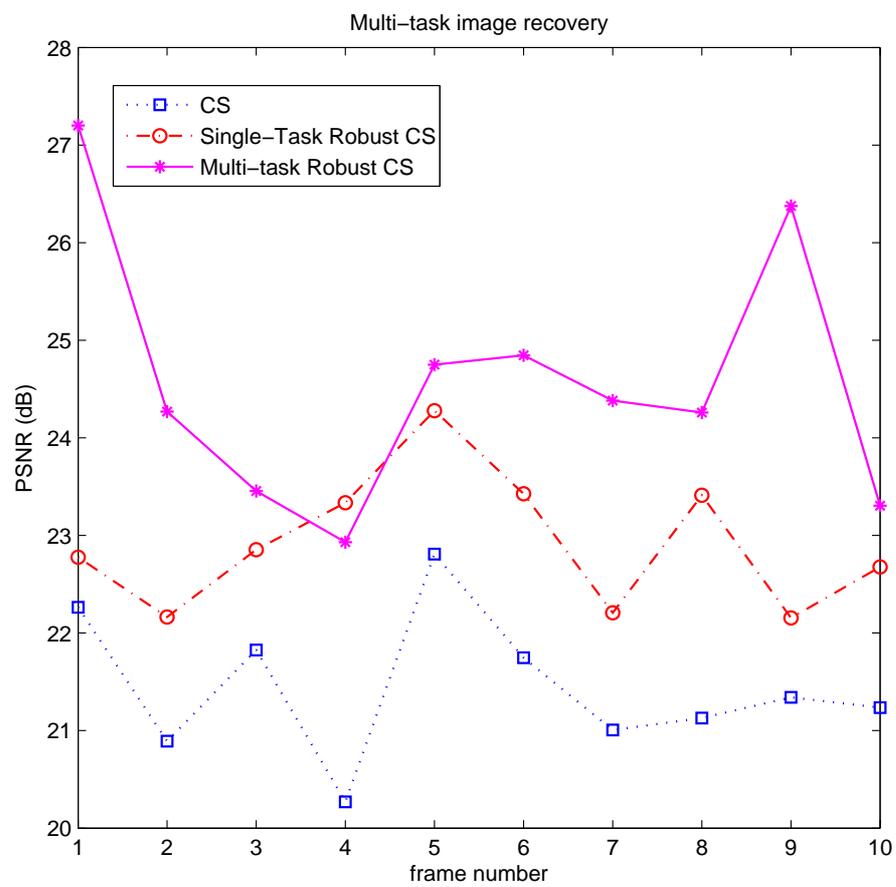}
	\caption{\label{FIG_MT_PSNR} PSNR Comparison of CS, robust CS, and multi-task (MT) robust CS}
\end{figure*}

The settings for the recovery are the same as previous experiments. For robust CS, we select the ADMM algorithm, and similarly for multi-task robust CS we also select the corresponding multi-task ADMM algorithm. The first 4 recovered images are shown in Fig. \ref{FIG_MT_RECOVERY}: the second row shows CS recovery, the third row shows robust CS recovery, and finally the last row shows multi-task robust CS recovery. The actual PSNRs for every frame are shown on Fig. \ref{FIG_MT_PSNR}. Here, we observe clearly that, on average, the multi-task robust CS formulation does provide a significant improvement over the robust CS formulation, both of which outperform CS recovery considerably.

\section{Conclusion}
We have presented more computationally efficient and extendable approaches to the recently proposed robust CS algorithm. We have also extended robust CS formulation in a number of ways, including affine constraints, $\ell_1$-loss function, and multi-task formulation. For improving computational efficiency of robust CS, we found that the (generalized) ADMM robust CS algorithm is the best, then followed by the FISTA robust CS algorithm. We also found that imposing affine constraint can provide improvement, though slightly. The striking result is that $\ell_1$ loss formulation for robust CS seems to offer considerable gain over the Huber's loss formulation, despite the fact that its convergence seems slower. Finally, in the case where one needs to robustly recover a sequence of compressed images, the multi-task formulation is proved to provide additional advantages in terms of both PSNR output and computational speed.
	

\section*{Appendices}
\subsection*{Proof of Lemma \ref{LEMMA_L_SUM}}
We start from the definition of the Lipchitz constant as a term such as
	\begin{eqnarray}
	 	\sup_{x_1,x_2\in\mathcal{X}} |f(x_1)-f(x_2)| \leq L_f|x_1-x_2|.
	\end{eqnarray}
As there are two possible scenarios $x_1,x_2\in\mathcal{X}_1$, $x_1,x_2\in\mathcal{X}_2$, and $x_1\in\mathcal{X}_1,x_2\in\mathcal{X}_2$ and from the definition of $L_g$ and $L_h$, we immediately have
	\begin{eqnarray}
	\label{EQU_SUP_LF}
	 L_f \leq \max\{L_g,L_h,L_{12}\},
	\end{eqnarray}
where $L_{12}$ is defined as the minimum constant such that
	\begin{eqnarray}
	 	\sup_{x_1\in\mathcal{X}_1,x_2\in\mathcal{X}_2} |g(x_1)-h(x_2)| \leq L_{12}|x_1-x_2|.
	\end{eqnarray}
Let $\mathcal{X}_3 = \mathcal{X}_1\cap\mathcal{X}_2$. For arbitrary $x_1\in\mathcal{X}_1$ and $x_2\in\mathcal{X}_2$ we construct $x_3\in\mathcal{X}_3$ such that it is a convex combination of $x_1$ and $x_2$, so that $|x_2-x_3|\leq |x_1-x_2|$ and $|x_1-x_3|\leq |x_1-x_2|$. Then  using triangle inequalities and definitions of $L_g$ and $L_h$, we have
	\begin{eqnarray}
	 	\sup |g(x_1)-h(x_2)| & = & \sup |g(x_1)-g(x_3)+g(x_3)+h(x_2)| \nonumber \\
		&= & \sup |g(x_1)-g(x_3)+h(x_3)-h(x_2)| \nonumber \\
		& \leq & \sup |g(x_1)-g(x_3)|+|h(x_3)-h(x_2)| \nonumber \\
		&\leq & \sup |g(x_1)-g(x_3)| + \sup |h(x_2)-h(x_3)| \nonumber \\
		&\leq & L_g|x_1-x_3| + L_h|x_2-x_3|		\nonumber \\
		&\leq & L_g|x_1-x_2| + L_h|x_2-x_1|		\nonumber \\
		&\leq & (L_g+ L_h)|x_1-x_2|. \label{EQU_SUP12}		
	\end{eqnarray}
The proof immediate follows from (\ref{EQU_SUP_LF}) and (\ref{EQU_SUP12}).

\bibliographystyle{plain}

\end{document}